\documentclass[showpacs,preprintnumbers,amsmath,amssymb,aps,prl,twocolumn,superscriptaddress,10pt]{revtex4-1}

\usepackage{amsfonts}
\usepackage[pdftex]{graphicx}
\usepackage{color}
\usepackage{bm}
\newcommand{\eqnref}[1]{Eqn.~\ref{#1}}
\newcommand{\figref}[1]{Fig.~\ref{#1}}
\newcommand{\e}[1]{\text{e}^{#1}}

\renewcommand{\vec}[1]{\mathbf{#1}}
\newcommand{\punc}[1]{\,#1}
\newcommand{\diffd}{\text{d}}
\newcommand{\neweqnline}{\nonumber\\}

\begin{document}
\title{Fluctuation-induced pair density wave in itinerant ferromagnets}
\author{G.J.~Conduit}
\email{gjc29@cam.ac.uk}
\affiliation{Cavendish Laboratory, University of Cambridge, CB3 0WA, UK}
\author{C.J.~Pedder}
\affiliation{London Centre for Nanotechnology,
 University College London, 17-19 Gordon St, London, WC1H 0AH, UK}
\author{A.G.~Green}
\email{andrew.green@ucl.ac.uk}
\affiliation{London Centre for Nanotechnology,
 University College London, 17-19 Gordon St, London, WC1H 0AH, UK}

\begin{abstract}
  Magnetic fluctuations near to quantum criticality can have profound
  effects.  They lead to characteristic scaling at high temperature which
  may ultimately give way to a reconstruction of the phase diagram and the
  formation of new phases at low temperatures. The ferromagnet UGe$_2$ is
  unstable to p-wave superconducting order -- an effect presaged by the
  superfluidity in He$^3$ -- whereas in CeFePO fluctuations drive the
  formation of spiral magnetic order. Here we develop a general quantum
  order-by-disorder description of these systems that encompasses both of
  these instabilities within a unified framework. This allows us to demonstrate
  that in fact these instabilities intertwine to form a new phase, a pair
  density wave.
\end{abstract}


\maketitle

\section{Introduction}

The physics of metallic systems tuned near to continuous, zero-temperature
phase transitions has provided a rich vein of experimental surprises and
theoretical
insights~\cite{RevModPhys.69.315,ColemanSchofieldNature2005,Sachdev1999}.
Fluctuations in such systems drive new critical behavior that inextricably
involves quantum dynamics, described by Hertz~\cite{Hertz1976} as quantum
criticality.  Over the past decade, it has become apparent that quantum
fluctuations have an even more profound effect on the magnetic phase
diagram~\cite{Belitz:1997cr,Belitz99,Efremov:2008oq,Maslov:2009kl,
  Rech:2006nx,Kirkpatrick:2012bs}. A breakdown of the standard
Moriya-Hertz-Millis~\cite{Moriya,Hertz1976,Millis:1993uq} model of itinerant
magnetic criticality occurs as the quantum critical point becomes occluded
by new fluctuation-driven phases. Magnetic fluctuations transverse to the
polarization drive many materials to display a first order ferromagnetic
transition~\cite{Belitz12}, spatially modulated magnetic order -- for
example CeFePO~\cite{Lausberg12i}, NbFe$_2$~\cite{Tompsett2010},
Yb(Rh$_{0.73}$Co$_{0.27}$)$_{2}$Si$_2$~\cite{Lausberg12ii} and possibly
Sr$_3$Ru$_2$O$_7$~\cite{Borzi07,Berridge2009}, or higher angular momentum
particle-hole pairing~\cite{Chubukov:2009fv,Karahasanovic:2012hc}. Moreover,
in materials such as UGe$_{2}$~\cite{Saxena00,Huxley01,Watanabe02},
URhGe~\cite{Aoki01}, and UCoGe~\cite{Huy07,Gasparini:2010} longitudinal
magnetic fluctuations can drive p-wave Cooper pairing in a similar manner to
He$^3$ superfluidity~\cite{Balian:1963ve,Anderson:1973zr,Brinkman:1974}.

\begin{figure}[!ht]
\includegraphics[width=1.\linewidth]{{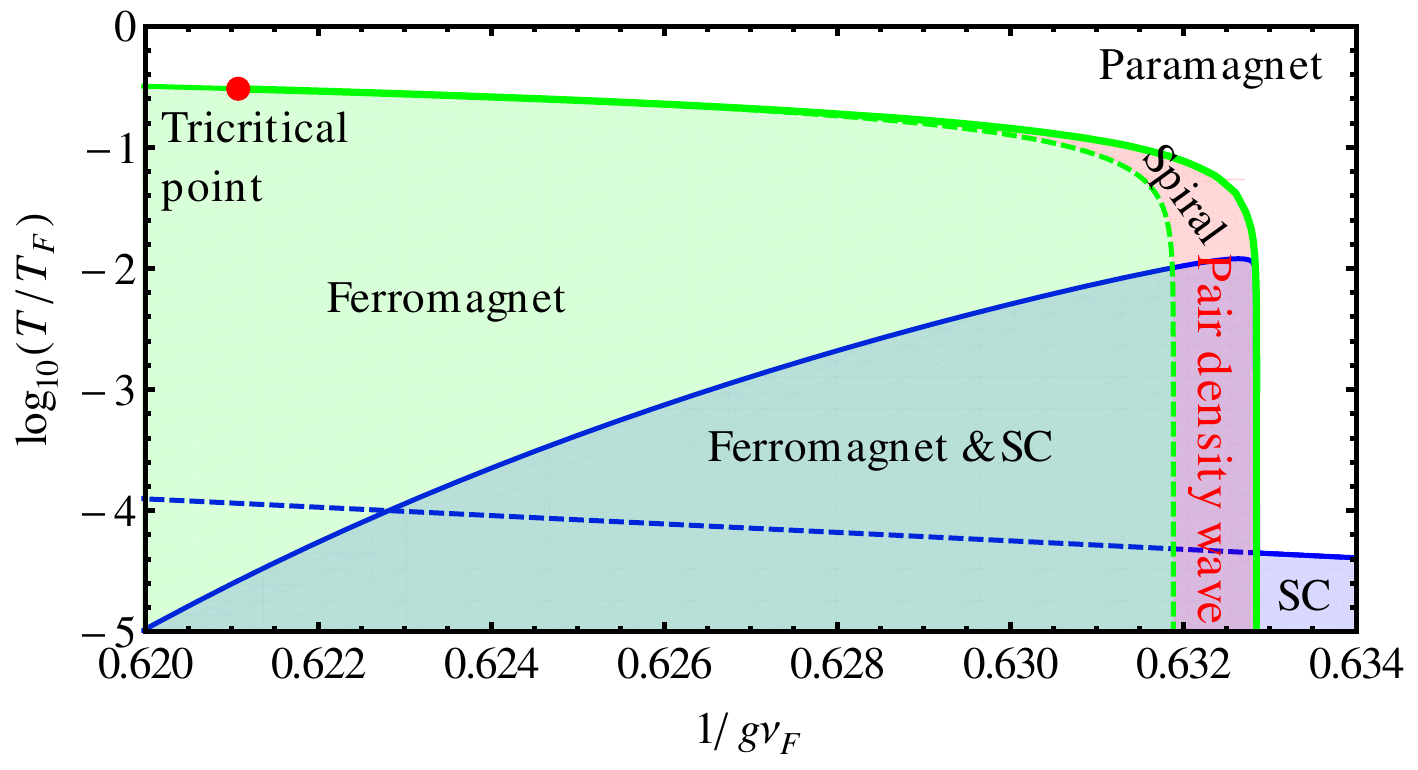}}
\caption{(Color online) The phase diagram of Hamiltonian,
  \eqnref{Hamiltonian} as a function of temperature and interaction
  strength. Our new phase, the pair density wave, is highlighted by the red
  text. $g$ is the interaction strength, $\nu_{\text{F}}$ is the density of
  states at the Fermi surface, $T$ temperature, and $T_{\text{F}}$ the Fermi
  temperature. \emph{The magnetic background} to this phase diagram is as
  determined in Ref.~\cite{Conduit:2009dq}, allowing for a resummation of
  the leading divergent contributions to the free energy. At high
  temperatures, there is a continuous transition from ferromagnet to
  paramagnet. The temperature of this reduces with decreasing interaction
  strength until the red tricritical point. At this point, there is a
  Lifshitz transition from the ferromagnet into a magnetic spiral (green
  dotted line). Finally, the spiral magnetic order gives way to
  paramagnetism in a first order transition.  \emph{Superconductivity} in
  the absence of magnetic order occurs below the blue dotted line. This
  transition temperature rises dramatically in the presence of magnetic
  order due to mode-mode coupling.  Where superconductivity overlaps the
  spiral magnetic order a Larkin-Ovchinnikov-Fulde-Ferrell-like pair density
  wave state is formed, highlighted by the red text.}
\label{fig:PhaseDiagram}
\end{figure}

In this Letter we study the minimal model of freely dispersing electrons
interacting through a contact repulsion.  We develop a general quantum
order-by-disorder approach to calculate the effects of fluctuations on the
free energy, leading to the phase diagram in \figref{fig:PhaseDiagram}. Near
to the itinerant ferromagnetic quantum critical point the electron gas forms
a previously unknown phase, a pair density wave where the magnetic order
forms a spiral modulation coexistent with a
Larkin-Ovchinnikov-Fulde-Ferrell-like spatial modulation of the d-vector of a
p-wave superconductor~\cite{larkin:1964zz,Fulde:1964fu,Neupert2011}.

The essence of our order-by-disorder approach is to expand the free energy
in fluctuations about some putative order. The order modifies the electron
dispersion and so the energy of the quantum fluctuations.  This offers an
alternative, physical view of the non-analytic corrections that overturn
Moriya-Hertz-Millis theory leading to the opportunity to undergo a first
order
transition~\cite{Belitz:1997cr,Belitz99,Duine2005,Conduit:2009bh,Belitz12},
spatially modulated magnetism~\cite{Conduit:2009dq,Kruger:2012tg}, a nematic
phase~\cite{Karahasanovic:2012hc}, and p-wave
superconductivity~\cite{Fay:1980kx,Mathur1998,Roussev:2001ys,
  Kirkpatrick:2001vn,Wang:2001kl}.  Near to the quantum critical point the
fluctuations self-consistently stabilize the ordered state. This mechanism
is similar in spirit, but subtly different in details, to several others:
the Coleman-Weinberg mechanism of field theory~\cite{Coleman:1973fk},
quantum order-by-disorder~\cite{OrderbyDisorder}, and the Casimir
effect~\cite{casimir1948attraction}. From the latter perspective, the Fermi
surface may be viewed as providing boundary conditions for quantum
fluctuations through Pauli exclusion. Modifying the Fermi surface by the
introduction of order modifies the spectrum of the fluctuations. This
perspective emphasizes the central role played by the underlying fermionic
statistics (which is absent in the conventional application of these
ideas). However, we have chosen to use the term quantum order-by-disorder to
be closest in spirit to the conventional usage.

In this Letter we first revisit the quantum-by-disorder approach
demonstrating how transverse magnetic fluctuations can generate a spatially
modulated magnetic phase and extend the analytical formalism to apply near
zero temperature. Secondly, we introduce a superconducting instability
driven by longitudinal spin fluctuations into the formalism and elucidate
the phase diagram. Finally, we discuss the consequences of the intertwined
spiral and superconducting instabilities for experimental systems.

\section{Quantum order-by-disorder in the itinerant ferromagnet}

We first review the order-by-disorder approach to address the magnetic phase
diagram ignoring any possible superconducting instabilities. Our starting
point is a Hamiltonian of freely dispersing electrons with
$\epsilon_{\vec{k}}=k^2/2$ interacting through a contact repulsion $g$
reflecting the screened Coulomb interaction present in metals
\begin{align} {\cal H}= \underbrace{\sum_{\vec{k},\sigma} \epsilon_{\bf
      k}\hat n_{{\bf k}}^{\sigma}}_{{\cal H}_{0}}
  +\underbrace{g\int\diffd{\bf r}\; \hat n_{{\bf r}}^{\uparrow}\hat n_{{\bf
        r}}^{\downarrow}}_{{\cal H}_{\text{int}}}\punc{.}
 \label{Hamiltonian}
\end{align}
Here $\hat n_{\vec{k}}^{\sigma}$ is the fermion occupancy in momentum space,
$\hat n_{\vec{r}}^{\sigma}$ in real space, spin
$\sigma\in\{\uparrow,\downarrow\}$, and we adopt atomic units with
$\hbar=m=1$ throughout. This model shows a continuous ferromagnetic
transition in mean-field, the zero-temperature terminus of which Hertz
termed the quantum critical point~\cite{Hertz1976}.  Hertz realized that
quantum fluctuations could modify the critical behavior at this quantum
critical point, however it has recently been shown that fluctuations can
have an even more dramatic effect, reconstructing the phase diagram
\cite{Belitz:1997cr,Efremov:2008oq,Maslov:2009kl,
  Rech:2006nx,Kirkpatrick:2012bs,Conduit:2009bh,Conduit:2009dq,Kruger:2012tg}.

The analysis of these effects proceeds as follows: first we consider a
background spiral magnetization,
$\vec{M}_{\vec{q}}=M(\cos\vec{q}\cdot\vec{r},\sin\vec{q}\cdot\vec{r},0)$
that modifies the mean-field electron dispersion $\epsilon_{\bf
  k}^{\sigma}(\vec{M}_{\bf q})=\epsilon_{\bf
  k}+\sigma\sqrt{(\vec{k}\cdot\vec{q})^2+(gM)^2}$.  The mean-field and
fluctuation corrections to the free energy are then calculated in the
presence of this background spiral. To leading order in the fluctuations,
the free energy is given by ${\cal F}={\cal F}_{\text{MF}}+{\cal
  F}_{\text{fluct}}$ with
\begin{align}
{\cal F}_{\text{MF}}
&=
-T\sum_{\sigma,{\bf k}}\ln[\e{(\epsilon^\sigma_{\bf k}(\vec{M}_{\bf q})-\mu)/T}+1]
+\frac{g}{2}M_{\bf q}^2\punc{,}
\end{align}
where $\mu$ is the chemical potential, $T$ temperature, and
\begin{align}
{\cal F}_{\text{fluct}}
&=
\!g^{2}\!\!\!\!\!\!\!\!\!
\sum_{\renewcommand*{\arraystretch}{0.6}
\begin{array}{c}
\scriptscriptstyle{{\bf k}_1\!+\!{\bf k}_2\!=\!{\bf k}_3\!+\!{\bf k}_4}\\
\scriptscriptstyle{{\bf p}_1\!+\!{\bf p}_2\!=\!{\bf p}_3\!+\!{\bf p}_4}\\
\scriptscriptstyle s
\end{array}
}\!\!\!\!\!\!\!\!\!
\frac{\langle
c^{\dagger}_{\vec{k}_{\!1}\!\uparrow}c^{\dagger}_{\vec{k}_{\!2}\!\downarrow}
c_{\vec{k}_{\!3}\!\downarrow}c_{\vec{k}_{\!4}\!\uparrow}
|s\rangle\!\langle s|
c^{\dagger}_{\vec{p}_{\!1}\!\uparrow}c^{\dagger}_{\vec{p}_{\!2}\!\downarrow}
c_{\vec{p}_{\!3}\!\downarrow}c_{\vec{p}_{\!4}\!\uparrow}
\rangle}
{\epsilon^\uparrow_{{\bf k}_1}+\epsilon^\downarrow_{{\bf k}_2}-\epsilon^\uparrow_{{\bf
      k}_3}-\epsilon^\downarrow_{{\bf k}_4}}
\punc{.}
\label{FreeEnergy1}
\end{align}
The expectation is taken in the thermal ensemble and the states $|s\rangle$
are virtual intermediate states generated by the action of the Hamiltonian.
The fluctuation contribution to the free energy is then calculated following
the prescription of Refs.~\cite{Duine2005,Conduit:2009bh}, with the
result~\cite{Regularization}
\begin{align}
{\cal F}_{\text{fluct}}
&=
-2g^2\!\sum_{\renewcommand*{\arraystretch}{0.6}
\begin{array}{c}
\scriptscriptstyle{{\bf k}_1+{\bf k}_2}\\
\scriptscriptstyle{={\bf k}_3+{\bf k}_4}
\end{array}}
\!\frac{n^\uparrow_{{\bf k}_1}n^\downarrow_{{\bf k}_2}
[n^\uparrow_{{\bf k}_3}+n^\downarrow_{{\bf k}_4}]}{
\epsilon^\uparrow_{{\bf k}_1}+\epsilon^\downarrow_{{\bf k}_2}-\epsilon^\uparrow_{{\bf k}_3}-\epsilon^\downarrow_{{\bf k}_4}
}
\punc{.}
\label{Fluctuations2}
\end{align}
Ref.~\cite{Conduit:2009dq} evaluated the ${\cal F}_{\text{fluct}}$
numerically at each value of $\vec{M}_{\bf q}$.  This revealed that the
low-energy phase space of fluctuations is enhanced by uniform or spiral
magnetic order. Good agreement with variational Monte Carlo
calculations~\cite{Conduit:2009dq} encourages us to develop a more amenable
Ginzburg-Landau expansion near to the tricritical
point~\cite{Karahasanovic:2012hc}. The free energy is a
functional of the mean-field dispersion in the presence of the modulated
order parameter, $ {\cal F}(M_{\bf q} ) \equiv {\cal F}[\epsilon_{\bf
  k}(\vec{M}_{\bf q})] = \langle{\cal F}[M^2+\theta^2q^2] \rangle -
\langle{\cal F}[\theta^2q^2] \rangle$ where $\theta = {\bf k}\cdot{\bf
  q}/|{\bf k}||{\bf q}|$ and $\langle\cdots\rangle$ indicates an angular
average. In a Taylor expansion of the free energy in $M^2$ and $q^2$, terms
are then linked by geometric factors stemming from the angular average
\begin{align}
{\cal F}
\!=\!\left(\!\alpha\!+\!\frac{2}{3}\beta q^2\!+\!\frac{2}{5}\gamma q^4\!\right)\!M^2
\!+\!(\beta\!+\!\gamma q^2) M^4\!+\!\gamma M^6\!\punc{,}
 \label{eqn:TaylorExpansion}
\end{align}
so the expansion contains only three coefficients $\alpha$, $\beta$, and
$\gamma$ that are functions of the interaction strength $g$ and the
temperature $T$. Reconstruction of the phase diagram is driven by a $\log
T$ contribution to the $M^4$ coefficient $\beta$.  The analytical form of these
parameters is known near to the tricritical
point~\cite{Karahasanovic:2012hc}.

We now seek a superconducting instability that occurs at very low
temperatures away from the tricritical point. In order to access this, we
must extend the region of validity of our expansion to lower temperatures
away from the vicinity of the tricritical point. The leading fluctuation
contribution, $M^4\ln T$ is accompanied by higher order terms
proportional to $M^6/T^2$, $M^8/T^4$ {\it etc}. These occur with pre-factors
allowing them to be re-summed exactly to give the following
contribution to the free energy
\begin{align}
\frac{\nu_{\text{F}}(g\nu_{\text{F}})^2}{6}\frac{(gM)^4}{\mu^2}
\ln\left[\frac{(gM)^2+(\pi T/4\e{\cal C})^2}{4 \mu^2}\right]
\punc{,}
\end{align}
where $\nu_{\text{F}}$ is the density of states at the Fermi surface and
${\cal C}$ is the Euler constant. This is consistent with that found in
previous studies~\cite{Conduit:2009dq,Belitz:1997cr,Maslov:2009kl}.  In
practice, sub-leading terms that are less divergent in $T$ have an important
effect upon the position of the phase boundaries at low temperatures away
from the tricritical point.  We find that they enhance the region of the
phase diagram occupied by the spiral phase making it consistent with the
broader region seen in variation Monte Carlo
calculations~\cite{Conduit:2009dq}. Unfortunately, re-summation of these
sub-leading divergences does not yield a simple analytical form and we have
not included them here.

\section{Superconductivity in quantum order-by-disorder}

We now wish to incorporate the possibility of Cooper pairing. We focus upon
A1 type pairing of spin-up electrons into a p-wave
state~\cite{Mackenzie:2003qa}, as it is the most favorable given the
background magnetic order. Other pairing symmetries can be treated
similarly.

The superconducting order overlaps with the fluctuation-induced spiral
modulation so we must construct superconducting order that can be treated
self-consistently in the order-by-disorder scheme. The natural choice is a
p-wave state in the basis that is diagonal in the presence of the background
spiral, {\it i.e.} in the basis where electron spins are polarized locally
parallel or anti-parallel with the background spiral order. This leads to a
state in which the d-vector of the p-wave superconductor rotates in concert
with the magnetization. The state may be referred to as a pair density wave
as it involves modulation of both magnetic and superconducting order, or
alternatively Larkin-Ovchinnikov-Fulde-Ferrell-like since it involves a
modulation of the superconducting order parameter, but in its SU(2) rather
than U(1) sector.

As it stands, the quantum order by disorder procedure requires that the
emergent order changes the mean-field dispersion. In the case of a contact
repulsion, superconductivity does not have this effect. One way around this
(used to study the possibility of nematic order in
\cite{Karahasanovic:2012hc}) is to include a field conjugate to the
superconducting order. After calculating the appropriate free energy and its
fluctuation corrections, a Legendre transformation recovers the Gibbs free
energy. Here, we use an equivalent self-consistent approach. The general
scheme is as follows: i. after having first diagonalized the electron states
in the magnetic background, we add and subtract a term
\begin{align}
\sum_{\bf k}\left(\Delta_{\bf k}\tilde
c^\dagger_{-{\bf k} \uparrow}\tilde c^\dagger_{{\bf k}\uparrow}+c.c\right)
\end{align}
in the free energy, where $\tilde c^\dagger_{{\bf k} \uparrow}$ creates a
state with spin polarized parallel to the background magnetic order, which
may be spiral in certain regions of the phase diagram.  ii. The terms ${\cal
  H}_0 -\sum_{\bf k} \left( \Delta_{\bf k} \tilde c^\dagger \tilde c^\dagger
  + c.c \right) $ can be diagonalized by a Bogoliubov transformation.
iii. The contributions of the terms ${\cal H}_{\text{int}} +\sum_{\bf k} \left(
  \Delta_{\bf k} \tilde c^\dagger \tilde c^\dagger + c.c \right) $ can be
calculated perturbatively in the Bogoliubov basis.  For BCS theory, the
resulting free energy has the same gap equation as the conventional
treatment. As we now show, using this approach in the case of a
contact repulsion recovers spin-fluctuation mediated
pairing~\cite{Fay:1980kx,Mathur1998,Roussev:2001ys,Kirkpatrick:2001vn,Wang:2001kl}.

The superconducting order does not change the expectation of the contact
interaction and so only enters the mean-field free energy through the
kinetic term. Fluctuation corrections are obtained following the
prescription developed for the fluctuation driven spin spiral; the
fluctuation contributions to the free energy are calculated from
Eq.(\ref{FreeEnergy1}), though now in the diagonal Bogoliubov
basis. Overall, after expanding to order $|\Delta|^2$, we obtain the free
energy~\cite{Linearize}
\begin{widetext}
\begin{align}
&{\cal F}(M,\Delta)=
-T\sum_{{\bf k},\sigma}
\ln(\e{-{\xi_{{\bf k}\sigma}}/T}+1)
+\frac{gM^2}{2}
-2g^2\!
\sum_{\renewcommand*{\arraystretch}{0.6}
\begin{array}{c}
\scriptscriptstyle{{\bf k}_1+{\bf k}_2}\\
\scriptscriptstyle{={\bf k}_3+{\bf k}_4}
\end{array}
}\!
\frac{n^\uparrow_{{\bf k}_1}n^\downarrow_{{\bf k}_2}
[n^\uparrow_{{\bf k}_3}+n^\downarrow_{{\bf k}_4}]}{
\epsilon^\uparrow_{{\bf k}_1}
+\epsilon^\downarrow_{{\bf k}_2}
-\epsilon^\uparrow_{{\bf k}_3}
-\epsilon^\downarrow_{{\bf k}_4}}
\neweqnline
&
-\!\!\sum_{\bf k}\!\frac{|\Delta_{\bf k}|^2}{2\xi_{{\bf k},\!\uparrow}}
(2n^\uparrow_{{\bf k}}\!\!-\!1)
\!+\!g^2\!\sum_{{\bf k},{\bf q}}\!
\frac{\bar \Delta_{{\bf k}+{\bf q}}}{2\xi_{{\bf k}\!+\!{\bf q},\!\uparrow}}
(1\!-\!2n^\uparrow_{{\bf k}\!+\!{\bf q}})
\frac{\Delta_{{\bf k}}}{ 2 \xi_{{\bf k},\!\uparrow}}
(1\!-\!2n^\uparrow_{{\bf k}})
{\cal R}e\chi^{\downarrow\downarrow}
({\bf q},\epsilon^\uparrow_{{\bf k}\!+\!{\bf q}}\!\!-\!\epsilon^\uparrow_{{\bf k}})
\!\!+\!\!\sum_{{\bf k}}\!
\frac{|\Delta_{{\bf k}}|^2}{2\xi_{{\bf k},\!\uparrow} } 
(1\!-\!2n^\uparrow_{{\bf k}})
\partial_{\epsilon_{\bf k}}{\cal R}e\Sigma^\uparrow({\bf k},\!\epsilon_{{\bf k}})
\!\punc{,}
 \label{FreeEnergy2}
\end{align}
where $\xi_{{\bf k}\sigma}=\epsilon_{\bf k}^\sigma-\mu$. $\chi$ and
$\Sigma$ are the susceptibility and self-energy evaluated at finite
$M_{\vec{q}}$ and spiral vector $\vec{q}$, and are given by
\begin{align}
{\cal R}e\chi^{\downarrow\downarrow}({\bf q},\omega)\!=\!
\!\sum_{{\bf p}}\!
\frac{n^\downarrow_{{\bf p}}-n^\downarrow_{{\bf p}+{\bf q}}
}{\epsilon^\downarrow_{{\bf p}\!+\!{\bf q}}\!-\!\epsilon^\downarrow_{\bf p}\!-\!\omega}
\hbox{ and }
\partial_{\epsilon_{\bf k}}
\Sigma^{\uparrow}({\bf k},\epsilon_{\bf k})
\!=\!
-g^2
 \partial_{\epsilon_{\bf k}}
 \!\sum_{{\bf p},{\bf q}}\!
 \frac{
 n^\downarrow_{{\bf p}-{\bf q}}
 (1\!-\!n^\uparrow_{{\bf k}-{\bf q}} )
 (1\!-\!n^\downarrow_{{\bf p}} )
 \!+\!
 (1\!-\!n^\downarrow_{{\bf p}-{\bf q}})
 n^\uparrow_{{\bf k}-{\bf q}} 
 n^\downarrow_{{\bf p}} 
 \!-\!n^\downarrow_{{\bf p}-{\bf q}}
}{
 \epsilon^\uparrow_{{\bf k}}
 +\epsilon^\downarrow_{{\bf p}-{\bf q}}
 -\epsilon^\uparrow_{{\bf k}-{\bf q}}
 -\epsilon^\downarrow_{{\bf p}}}
\!\punc{.}
\end{align}
\end{widetext}
The first line of \eqnref{FreeEnergy2} that is independent of $\Delta$
is precisely the free energy of the ferromagnet derived earlier.  At finite
$\Delta$, the first term of the second line contains the mean field response
of the Fermi surface to Cooper pairing, and the subsequent terms arise from
the fluctuation corrections.

The gap equation resulting from this is given by exactly the same expression
as the Eliashberg equations obtained in spin-fluctuations theory.  The
resulting transition temperature for the pair density wave can be written
as~\cite{Fay:1980kx,Mathur1998,Roussev:2001ys,Kirkpatrick:2001vn,Wang:2001kl}
\begin{equation}
T_{\text{c}}\!=\!
 \frac{2 \mu^\uparrow\e{\cal C}}{\pi } 
\exp\!\left[-\frac{
\left(1- \partial_\epsilon {\cal R}e \Sigma^\uparrow ({\bf k}, \epsilon_{\bf k}) \right)
\langle \langle
\theta^2_{{\bf k}} 
\rangle \rangle
}{
\langle \langle
\theta_{{\bf k}+{\bf q}} 
\theta_{\bf k}
 {\cal R}e \chi^{\downarrow \downarrow}({\bf q},\epsilon_{{\bf k}\!+\!{\bf q}}\!-\!\epsilon_{{\bf k}})
\rangle \rangle 
}
\right]\!\!\punc{,}
\label{TransitionTemperature}
\end{equation}
where ${\cal C}$ is the Euler constant, and $\theta_{\bf k}$ is the angular
dependence of the superconducting order parameter~\cite{AngularAverages}.
\eqnref{TransitionTemperature} gives the transition temperature in
paramagnetic, ferromagnetic and spiral phases provided $\chi$ and $ \Sigma$
are calculated in the appropriate backgrounds.  We note that both the
fluctuation-induced pairing and field renormalization occur automatically in
this approach. The field renormalization/self-energy term written here
includes the effect of the regularization described above
\eqnref{FreeEnergy1}. Because of the intervention of phase reconstruction
before the quantum critical point is reached, we do not require that the
magnetic fluctuations take their critical
form~\cite{Fay:1980kx,Mathur1998,Roussev:2001ys,Kirkpatrick:2001vn,Wang:2001kl}.

The dependence of $\Sigma$ and $\chi$ upon the background magnetic order
through the mean-field dispersion, $\epsilon^\sigma_{\bf k}(M,{\bf
  q})=\epsilon_{\bf k} + \sigma \sqrt{ (g M)^2+ (\theta_{\bf k}\cdot{\bf
    q})^2 }$, induces cross terms between the superconducting and magnetic
order parameters and their gradients. Indeed such cross terms also appear in
the mean-field expansion of the bare Cooper susceptibility and ultimately in
the pre-factor of chemical potential in
Eq.(\ref{TransitionTemperature}). Terms arising from the fluctuation
corrections are precisely the mode-mode coupling terms included by
Kirkpatrick {\it et al.} previously~\cite{Kirkpatrick:2001vn}.

In practice, we find that although the gradient terms are crucial in
driving the pair density wave state, expanding the susceptibility $\chi$ and
self-energy $\Sigma$ in the spiral wave vector leads to sub-leading
corrections that do not change the superconducting transition temperature
appreciably. At small and large magnetization, we find:
\begin{align*} 
&g^2\langle\langle
\theta_{{\bf k}+{\bf q}} 
\theta_{\bf k}
 {\cal R}\chi^{\downarrow\downarrow}
({\bf q},\epsilon_{{\bf k}+{\bf q}}-\epsilon_{{\bf k}})
\rangle\rangle/\langle\langle\theta_{\bf k}^2\rangle\rangle
\neweqnline
&\approx\!
\begin{cases}
\frac{(g\nu_{\text{F}})^2}{5}
\left[2\ln2-1+
\frac{5}{3}\frac{gM}{ \mu}\left(7 -4 \ln 2 \right)\right]
&\hbox{Small }M\\
\frac{(g\nu_{\text{F}})^2}{2}
\left(1-\frac{gM}{\mu}\right)^{3/2}\!\!
\left(1+\frac{1}{3}\ln\left[2-2\frac{gM}{\mu}\right]\right)
&\hbox{Large }M
\end{cases}
\end{align*}
and
\begin{align*}
 \langle\langle\partial_{\epsilon_{\bf k}}\Sigma({\bf k},\epsilon_{\bf k})\rangle\rangle
\!\approx\!\begin{cases}
(g\nu_{\text{F}})^2\left[\left(1\!-\!\frac{gM}{\mu}\right)\!\ln2+\frac{1}{2}\right]
&\hbox{Small }M\\
\frac{(g\nu_{\text{F}})^2}{2^{3/2}} \sqrt{1-\frac{g M}{\mu}}
&\hbox{Large }M
\end{cases}
\end{align*}

\section{Phase diagram}

The phase diagram is shown in \figref{fig:PhaseDiagram}. With a first order
transition and spin spiral emerging at low temperatures, the phase diagram
is a refinement of that presented in Ref.~\cite{Conduit:2009dq}, the one
notable difference being the order of the transition from the paramagnet to
the spin spiral. In Ref.~\cite{Conduit:2009dq} only the lowest order term in
the free energy expansion was calculated, implicitly assuming a second order
paramagnet-spiral transition. Here we have extended the free energy
expansion to order $M^6$, which shows that the paramagnet-spiral transition
is in fact first order~\cite{Karahasanovic:2012hc}. Though this phase
diagram is consistent with variational Monte Carlo
results~\cite{Conduit:2009dq}, the spiral phase here occupies a smaller
range of interaction strengths due to the absence of sub-leading
divergences. However the topology of the phase diagram is identical so we
proceed with the amenable Ginzburg Landau formalism. As the
superconductivity is weak it may be laid over the magnetic phase diagram without
appreciable feedback.

Magnetic fluctuations drive Cooper pairing, so that even at $M=0$ a
superconducting phase is formed. When the superconductivity overlaps regions
of the phase diagram supporting magnetic order, there is a large increase in
the transition temperature because of the mode-mode coupling effects
described above and in Ref.~\cite{Kirkpatrick:2001vn}. The transition
temperature tails off exponentially to zero when the magnetization has
depleted the number of pairing (spin-down) electrons to zero thus sending
the susceptibility term to zero. In \figref{fig:PhaseDiagram}, the
superconducting transition temperature shows a first order jump upon passing
into the magnetically ordered phase, tracking the local magnetization. Where
the superconductivity overlaps the spiral phase, its d-vector is locked to
the local magnetic order forming a new phase, the pair density wave.

\section{Discussion}

The minimal Hamiltonian Eq.(\ref{Hamiltonian}) -- an electron gas with
contact repulsion -- delivers a remarkably rich phase diagram when all
possible fluctuations are accounted for.  Fluctuations drive the itinerant
ferromagnet near to several possible instabilities. Transverse
fluctuations can drive a first order ferromagnetic transition or a spin
spiral phase.  Longitudinal magnetic
fluctuations drive a p-wave superconducting instability.
These instabilities overlap forming a pair density
wave.  Comparison with variational Monte Carlo
calculations~\cite{Conduit:2009dq} shows that the quantum order-by-disorder
approach, as well as revealing the underlying physics in a transparent
manner, can accurately determine the phase boundaries.

However, the model may require modification in more realistic
settings. Firstly, unlike the local interaction used here, the screened
Coulomb interaction has finite range and is sampled at different momenta by
mean-field and fluctuation contributions to the free energy. The tricritical
point -- the lowest temperature at which a continuous ferromagnetic to
paramagnet transition is observed -- is commonly seen at temperatures
$\sim0.02T_{\text{F}}$~\cite{Borzi07} that are significantly lower than the
prediction of the minimal Hamiltonian
$\sim0.3T_{\text{F}}$. However, it has been shown both with variational Monte Carlo
and analytically that a finite ranged screened Coulomb interaction reduces
the tricritical temperature found in a quantum order-by-disorder calculation
to the realistic experimental range of
$\sim0.02T_{\text{F}}$~\cite{Conduit2013}. Secondly, real band structures,
with deviations from quadratic electron dispersions, shift the subtle
balance between the near multi-critical
phases~\cite{Karahasanovic:2012hc}. Finally, spin-orbit induced magnetic
anisotropy can suppress transverse magnetic fluctuations relative to the longitudinal
fluctuations so changing the balance between magnetic and
superconducting instabilities~\cite{Mathur1998},
and may ultimately invalidate the neglect of back-reaction of superconducting
order upon the magnetic order.

The necessity of these modifications to the model Hamiltonian makes direct
comparison of our results with experiment delicate. The types of order
predicted in our model calculations have all been seen in (or suggested by)
experiment. For example p-wave superconducting is found in
UGe$_{2}$~\cite{Saxena00,Huxley01,Watanabe02}, URhGe~\cite{Aoki01}, and
UCoGe~\cite{Huy07}. In the magnetic state UGe$_{2}$ displays
superconductivity below $T_{\text{c}}\approx500\text{mK}$, and no
superconductivity has been observed on the paramagnetic side. Theory
predicts that the peak superconducting temperature has a ratio of $200:1$
between the ferromagnetic and paramagnetic sides, meaning that on the
paramagnetic side UGe$_{2}$ would have a transition temperature at
$T_{\text{c}}\approx3\text{mK}$ that is too small to measure, thus
maintaining consistency with experiment~\cite{Kirkpatrick:2001vn}. Evidence
for spin spiral behavior, though less direct, is found in
CeFePO~\cite{Lausberg12i}, NbFe$_2$~\cite{Tompsett2010},
Yb(Rh$_{0.73}$Co$_{0.27}$)$_{2}$Si$_2$~\cite{Lausberg12ii} and possibly
Sr$_3$Ru$_2$O$_7$~\cite{Borzi07,Berridge2009}. Although evidence for the
coexistence of superconductivity and spiral magnetism has not yet been
found, it could be measured through a drop of electrical resistance coupled
with the emergence of spatial texture in a neutron scattering
study. Nevertheless, we take the occurrence of these ingredients as reason
for positive expectation.

The quantum order-by-disorder framework provides a unifying and physical
reformulation of diagrammatic approaches to particle-hole and
particle-particle instabilities. The Fermi surface provides boundary
conditions for fluctuations through Pauli exclusion. Modification of the
Fermi surface by the introduction of new order affects the spectrum of
fluctuations, and can be self-consistently stabilized by them. We have shown
that this drives the formation of a new phase, the pair density wave. This
new formalism guides future calculations in a direction that may be fruitful
in a variety of situations.

\emph{Acknowledgments:} We are grateful to Una Karahasanovic and Frank
Kr\"uger for useful discussions, and acknowledge Gonville \& Caius College
and the EPSRC (grant EP/I004831/1) for financial support.

\end{document}